\documentclass[a4paper,11pt]{article}
\usepackage{pos}
\usepackage{lineno}

\title{Imaging Atmospheric Cherenkov Telescopes pointing determination using the trajectories of the stars in the field of view.}
 \ShortTitle{Star tracking for pointing determination}

\author*[a]{Mykhailo Dalchenko}
\author[a]{Matthieu Heller}

\affiliation[a]{Université de Genève, DPNC,\\
 24, quai Ernest-Ansermet, Genève, Switzerland}

\onbehalf{on behalf of the CTA-LST Project} 

\emailAdd{mykhailo.dalchenko@unige.ch}

\abstract{We present a new approach to the pointing determination of Imaging Atmospheric Cherenkov Telescopes (IACTs). 
This method is universal and can be applied to any IACT with minor modifications. 
It uses the trajectories of the stars in the field of view of the IACT's main camera and requires neither dedicated auxiliary hardware nor a specific data taking mode. 
The method consists of two parts: firstly, we reconstruct individual star positions as a function of time, taking into account the point spread function of the telescope; 
secondly, we perform a simultaneous fit of all reconstructed star trajectories using the orthogonal distance regression method. 
The method does not assume any particular star trajectories, does not require a long integration time, and can be applied to any IACT observation mode. 
The performance of the method is assessed with commissioning data of the Large-Sized Telescope prototype (LST-1), showing the method's stability and remarkable pointing performance of the LST-1 telescope.}

\ConferenceLogo{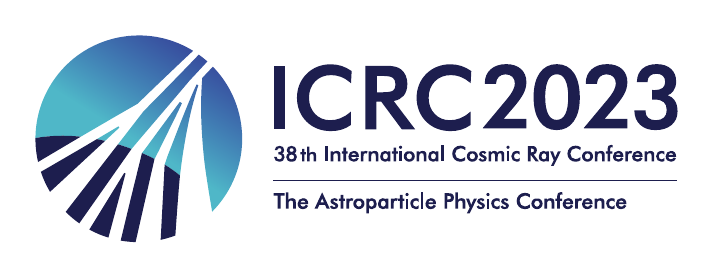}

\FullConference{%
38th International Cosmic Ray Conference (ICRC2023)\\
  26 July - 3 August, 2023\\
  Nagoya, Japan}

\begin{document}
\maketitle

\section{Introduction}
The next generation of Imaging Atmospheric Cherenkov Telescopes (IACTs) aims to achieve an exceptional angular resolution that enables detailed imaging of extended objects and the study of gamma-ray source morphology. The Cherenkov Telescope Array (CTA) is designed with the goal of attaining an angular resolution of $\leq1'$ at the analysis level~\citep{acharya:hal-01998995}.

However, the remarkable dimensions of modern IACTs, along with the substantial weight of their cameras, introduce mechanical deformations in the telescope structure. Factors such as wind and temperature variations further contribute to these deformations, potentially affecting the telescope's pointing accuracy. Therefore, continuous monitoring and correction of the telescope's pointing behavior are crucial to ensure the attainment of the required pointing accuracy.

Traditionally, the monitoring and correction of telescope pointing direction rely on dedicated auxiliary devices. For instance, the Large-Sized Telescope (LST) of CTA incorporates several pointing hardware components~\citep{Zaric:2019dv}:
\begin{itemize}
\item Starguider camera (SG): located at the center of the dish, the SG includes a CCD camera with a wide field of view (FoV) that captures the part of the Cherenkov camera with its reference LEDs and stars in the vicinity of the telescope's pointing direction. Analysis of its images allows us to express the position of the Cherenkov camera center in sky coordinates with a precision of approximately $5''$ at a frequency of around $1$\,Hz.
\item Camera Displacement Monitor (CDM): coupled to the SG camera, the CDM is a CMOS camera operating at approximately 10 frames per second. It measures the displacement of the Cherenkov camera's center with respect to the telescope's optical axis. 
\item Set of reference lasers, LEDs, and distance meters: these devices are used to determine the relative positions of the telescope structure, Cherenkov camera, and the aforementioned pointing hardware components.
\end{itemize}

In this work, we propose an analytical method that does not require specific hardware or dedicated data-taking. It exploits the rotational motion of off-axis stars in the telescope's field of view when the telescope tracks an object of interest. By reconstructing the trajectories of these stars, we can determine the telescope's pointing.
Similar approaches have been employed in previous experiments (see, e.g. \cite{kifune_1993}, \cite{thesis_braun_HESS}, and \cite{astri-astrometry}). These past applications rely on the photo-sensor currents analysis, followed by circular\footnote{Elliptical in case of ASTRI-Horn telescope} fit of the reconstructed star trajectories, requiring long integration periods to achieve the necessary precision for star position reconstruction. In contrast, our proposed method uses the physics data stream from the Cherenkov camera of the telescope, providing a higher data acquisition rate compared to the monitoring frequency of photo-sensor currents. Furthermore, our telescope's pointing reconstruction technique directly incorporates the temporal evolution of star positions in IACT camera images, allowing us to achieve a much higher pointing monitoring frequency, potentially exceeding 10\,Hz. By delivering frequently updated corrections to the telescope's pointing direction, we can account for deviations introduced by telescope structural deformations and environmental effects that can vary on a short timescale, on the order of minutes.

\section{Star Imaging with IACTs}

In star imaging with IACTs, there are three key factors that play a crucial role: the camera readout coupling, the telescope geometry, and its optical properties. These factors determine the ability to observe a continuous light signal and the distortion of point-like object images, such as stars.

\subsection{Impact of Telescope Optics on Star Image}
\label{sec:psf}

The Point Spread Function (PSF) represents the blurring effect occuring in real optical systems, causing a star to appear as a smeared-out image rather than a point-like object. In the case of the LST and other telescopes with parabolic optical systems composed of multiple facets, the PSF is a result of the convolution of facet aberrations. The PSF is affected by various factors, including imperfect mirror surfaces, facet alignment, and coma aberration. Although the facet alignment is continuously monitored and corrected in the LST through an active mirror control system, the coma aberration inherent to the parabolic system cannot be completely eliminated. In addition, star images are affected by defocus aberration due to IACTs being focused on air showers maximum at distances around 15\,km from the telescope.

To determine the PSF shape, a series of star simulations at different positions with respect to the pointing direction are performed using the \texttt{sim\_telarray}  software package~\citep{simtelarray}. The simulation model consists of LST-1 telescope parameters and an atmospheric model for the Roque de los Muchachos Observatory (ORM), where the LST-1 is situated. The coma aberration is accurately simulated, while the additional correction to the PSF due to imperfect mirror surfaces and facet alignment is extracted from PSF measurements performed with the specific hardware of the LST-1.

\begin{figure*}[h!]
	{\includegraphics[width=0.4\textwidth]{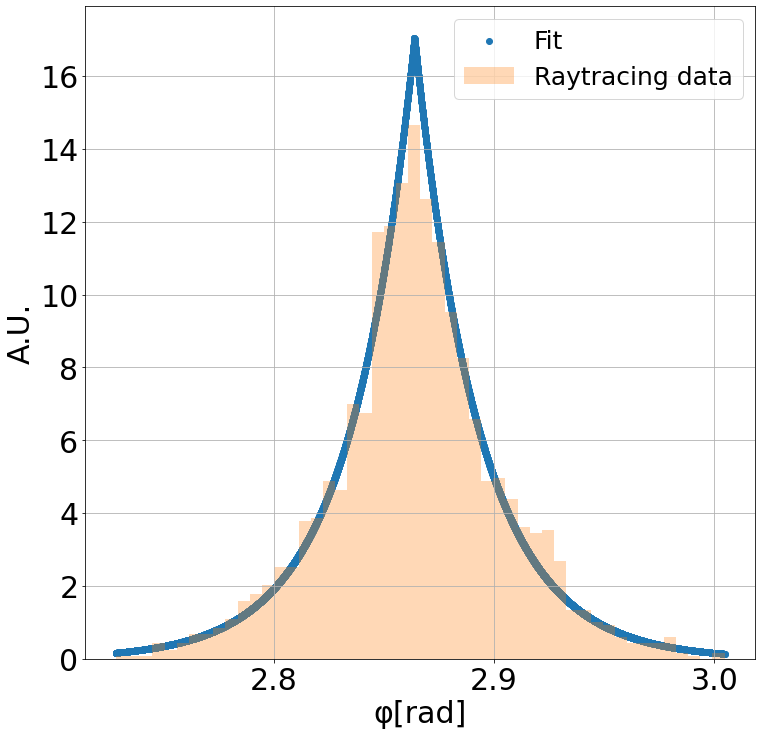}}
	\hfill
	{\includegraphics[width=0.4\textwidth]{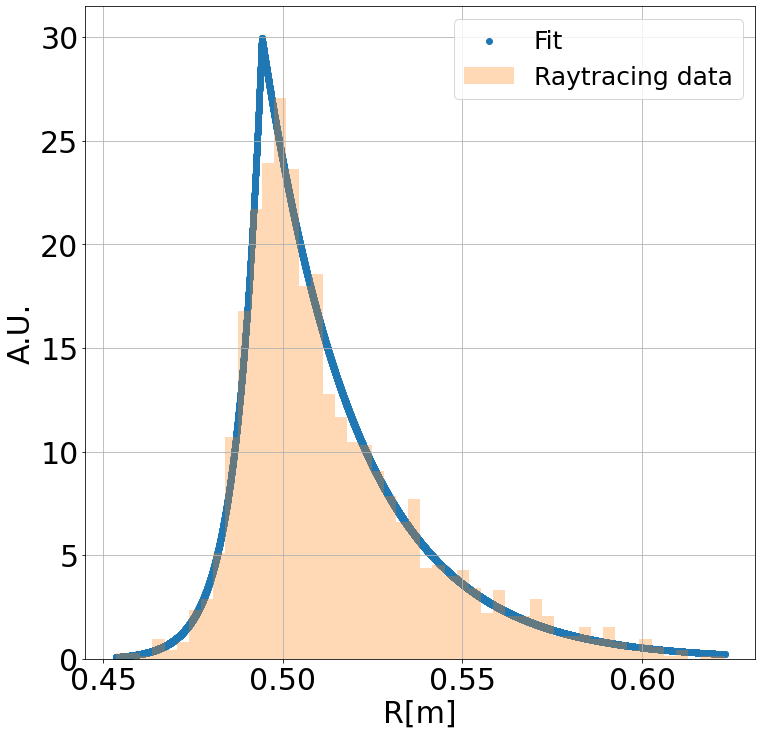}}
	\caption{PSF fit along the $\phi$ (left) and $r$ (right) polar coordinates for the LST-1.}
	\label{fig:psf_coma_pdf_fit}
\end{figure*}

The starlight distribution in the camera frame is analyzed in polar coordinates. The azimuthal~($\phi$) and radial~($r$) components of the distribution are well modeled using a symmetric and asymmetric Laplace functions respectively:

\begin{equation}
\begin{aligned}
	f_{\phi}(x) = \frac{1}{2S_\phi}e^{-|\frac{x-L}{S_\phi}|},\qquad\qquad
    f_{r}(x, K) = \begin{cases}\frac{1}{S_{r}(K+K^{-1})}e^{-K\frac{x-L}{S_{r}}}, x\ge L\\ \frac{1}{S_{r}(K+K^{-1})}e^{\frac{x-L}{KS_{r}}}, x < L\end{cases}
\end{aligned}
\end{equation}

The parameters of these functions, such as scale ($S_\phi$, $S_r$), asymmetry ($K$), and the maximum position ($L$), are determined by fitting the simulated data. Figure~\ref{fig:psf_coma_pdf_fit} illustrates an example of the PSF fit along $\phi$ and $r$ for one star simulation at a distance of $0.8^\circ$ from the camera center. The simulated data (histograms) show good agreement with the fitted PSF curves (solid lines) for both the angular and radial components.

The dependency of the PSF parameters on the polar angle is negligible due to the axial symmetry of the telescope. Therefore, the parameterization is solely based on the radial distance. 
By performing star simulations at various offsets with respect to the camera center and analyzing the corresponding PSF parameters, an analytic dependency of these parameters on the radial distance can be established:

\begin{equation}
    \begin{aligned}
        S_{r}(r) & = a_1 - a_2\,r + a_3\,r^2 + a_4\,r^3\\
        S_{\phi}(r) & = b_1\,\exp{(-b_2\,r)}+\frac{b_3}{b_3+r} \\
        K(r) & = 1 - c_1\,\tanh{(c_2\,r)}+c_3\,r \\
    \end{aligned}
    \label{eq:laplace_param}
\end{equation}

\subsection{Cleaning and Averaging Algorithm for Star Image Reconstruction}

In the case of the LST, which has AC-coupled readout electronics\footnote{AC-coupled readout electronics block the DC component of a signal, making the readout insensitive to changes in the baseline level.}, there are two options to obtain a star image in the photodetection plane:
\begin{itemize}
\item Using the photomultiplier anode current in each pixel: this method allows for direct observation of the star flux. However, the frequency at which the anode current is monitored is typically below 1 Hz. As a result, the rate at which the star position can be reconstructed and the pointing updated is naturally limited. It is important to note that anode current values are not included in the physics data stream and are considered auxiliary variables. Therefore, accessing these values requires the development of a specific data access interface.
\item Using the variance of the signal amplitude in each pixel: this method is universal and can be applied to any telescope's physics data stream without the need for custom data formats or access interfaces, provided that the full waveform is available. This technique is not strictly limited to AC-coupled readout electronics and can apply to telescopes equipped with DC-coupled readout electronics. In the case of DC-coupled electronics, the reconstruction performance can be further improved by directly using the DC baseline level.
The advantage of using the signal amplitude variance method is its compatibility with standard data formats and access interfaces, making it a more versatile option for star imaging.
\end{itemize}
Based on these considerations, we opt to explore star imaging through the analysis of signal amplitude variances.

In order to achieve accurate star reconstruction, the variance image of the camera, which is a snapshot of the camera with the variance calculated for each pixel, needs to be properly cleaned. This cleaning process involves removing the effects of extensive air showers (EAS) and the night sky background (NSB) photons.
The cleaning and averaging algorithm for star image reconstruction follows the steps outlined below:

\begin{enumerate}
    \item Prepare a calibrated events stream: the algorithm starts with a calibrated events stream, where the pixel waveform amplitude is provided in photoelectrons (p.e.). For simulated data, calibration is straightforward, while for observation data, dedicated software provided by the LST-1 collaboration is used \citep{lstchain}.
    \item Clean the events from EAS contamination: the \texttt{LocalPeakWindowSum} charge extraction algorithm \citep{ctapipe} is applied to each pixel to produce a reconstructed charge image. The integration window shift and width are set to 4 and 8\,ns, respectively, which are the default values for LST-1. The standard LST image reconstruction tools \citep{ctapipe} are then used to determine the pixels affected by EAS. These tools utilize default cleaning parameters such as picture threshold (7 p.e.), boundary threshold (5 p.e.), and no isolated pixels.
    \item Replace EAS-affected pixel variances: the variances of the pixels affected by EAS are replaced with the average pixel variance value of the complete camera for that event. This average pixel variance corresponds to the NSB level. To calculate this average pixel variance, the following criteria are applied:
    \begin{itemize}
        \item The pixel is not affected by EAS photons.
        \item The pixel is not in the vicinity of the expected reconstructed star position.
        \item The pixel is switched on with regular gain settings.
    \end{itemize}
    \item Compute the clean average variance image: using $N$ consecutive events, compute the average variance image and subtract the NSB contribution. The average NSB contamination is computed considering the pixels, fulfilling the criteria from the previous step.
\end{enumerate}

To find the optimal averaging window size we performed a simulation of 300 consecutive events\footnote{Diffuse proton-induced EAS with primary particle energy distributed between 10\,GeV and 100\,TeV with spectral index equal to -2.0} with several typical stars, using the \texttt{sim\_telarray}. A stable integrated variance is observed after averaging over 200 events, which we use in the following.

\subsection{Star Position Reconstruction}

Once the cleaned and averaged variance image is obtained, the next step is to perform star position reconstruction and estimate the associated uncertainties. Using the PSF parameterization, described in Sec.~\ref{sec:psf}, we identify the clusters of pixels observing the star photons according to the following rule: a pixel is included in the cluster, if the PSF integrated over its area exceeds $0.1\%$ of the total PSF integral. The star is considered detected if at least one pixel from its cluster has its variance surpassing three standard deviations of the NSB-only signal. Once the clusters are identified, the star positions are calculated by averaging the positions of all pixels within the clusters, weighted by their variance values. The reconstructed position uncertainty is computed as the covariance of the pixel's center coordinates, using the PSF values integrated per pixel as weighting factors. The simulations show that the position reconstruction achieves an accuracy of $20''$ and a precision of $25''$ when reconstructing a star located $1^\circ$ away from the optical axis of the telescope.

\section{Star Trajectory Fitting}

The expected trajectory of each star in the telescope's camera frame, denoted as $\vec{x}_i$, is represented as an implicit function of time~($t$), star position in the International Celestial Reference System (ICRS) frame ($\vec{c}_i$) and the telescope's pointing in the local Altitude-Azimuth (AltAz) frame ($\vec{p}(t)$):

\begin{equation}
\vec{x}_i = X(t, \vec{c}_i, \vec{p}(t))
\label{eq:coord}
\end{equation}

The coordinate frames are defined as follows:
\begin{itemize}
\item ICRS, aligned close to the mean equator and dynamical equinox of J2000.0~\citep{iau_astro_ref_systems}.
\item Local horizontal frame in the Altitude-Azimuth system with respect to the WGS84~\citep{nima:2000} ellipsoid (AltAz).
\item Local camera coordinate frame. The camera frame is a 2D Cartesian frame that describes the position of objects in the focal plane of the telescope. Starting at the horizon, the telescope is pointed to the magnetic North in azimuth and then up to the zenith. Now, abscissa~($x$) points North, and ordinate~($y$) points West.
\end{itemize}

Ideally, when the telescope is tracking a celestial object, the pointing direction in the ICRS frame remains constant with the RA and Dec coordinates of the tracked source. However, when transformed into the AltAz frame, the pointing direction becomes time-dependent. Additionally, we introduce a time-dependent pointing displacement $\Delta \vec{p}$ as a correction to the pointing direction $\vec{p}$ reported by the drive system. This displacement accounts for structural deformations of the telescope that were described previously.

While a three-dimensional representation of the star trajectory is possible, allowing for rotations, tilting of the Cherenkov camera, or changes in the telescope's focal length, we focus on a two-dimensional representation. This approximation is based on the fact that the main reason for the telescope's mispointing is the planar displacement of the Cherenkov camera caused by the bending of the telescope structure.

We apply Orthogonal Distance Regression (ODR)~\citep{odr} to reconstruct the pointing deviation observed by the telescope from the nominal position provided by the drive system. It minimizes the distance between the star positions provided by the catalog and the reconstructed star positions, yielding the correction to the nominal telescope pointing.
One advantage of using the ODR method is that it does not require assuming a specific trajectory shape for the stars, such as a circle or ellipse. This flexibility allows for its application to various telescopes, including satellite-based telescopes where stars may not follow circular trajectories. Another key factor in choosing the ODR method is its ability to handle uncertainties associated with all dependent variables, namely the star position coordinates.

\begin{figure}[t!]
	\centering\includegraphics[width=0.8\textwidth]{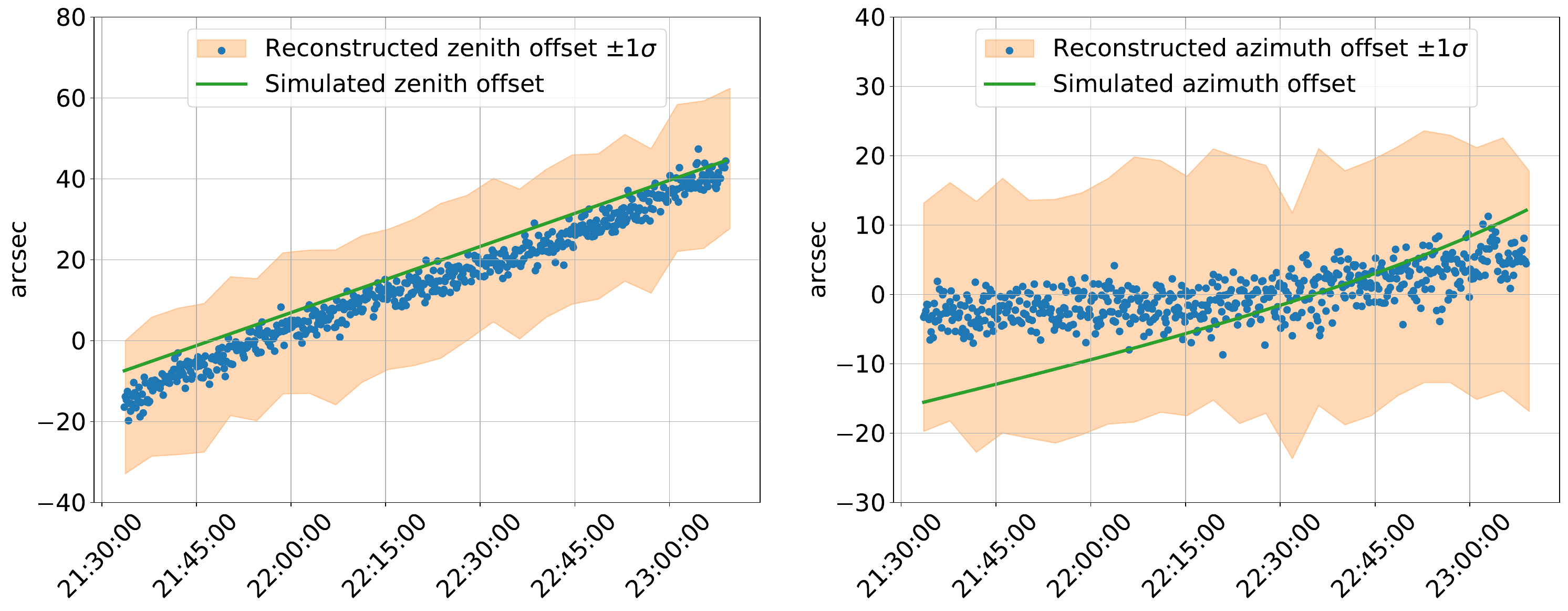}
	\caption{Reconstructed zenith distance (left) and azimuth (right) components of mispointing as a function of time for the simulation with the artificially introduced variable mispointing.}
	\label{fig:alt_az_mispointing_vs_time}
\end{figure}

\section{Results}

We evaluate the performance of the star tracking method on simulated and observed LST-1 data. For the simulations, the inputs were tuned to reflect real observation conditions with the LST-1 telescope, including the observed star field and latest studies on its characteristics. EAS were not simulated as their impact on star position reconstruction is negligible after cleaning and averaging procedures. A progressive telescope mispointing amounting to $2.4''$ per each degree of the telescope motion in zenith and azimuth direction was added to the simulations in order to reflect the potential pointing deviation under real data taking conditions.

\begin{figure}[!b]
	\centering\includegraphics[width=0.8\textwidth]{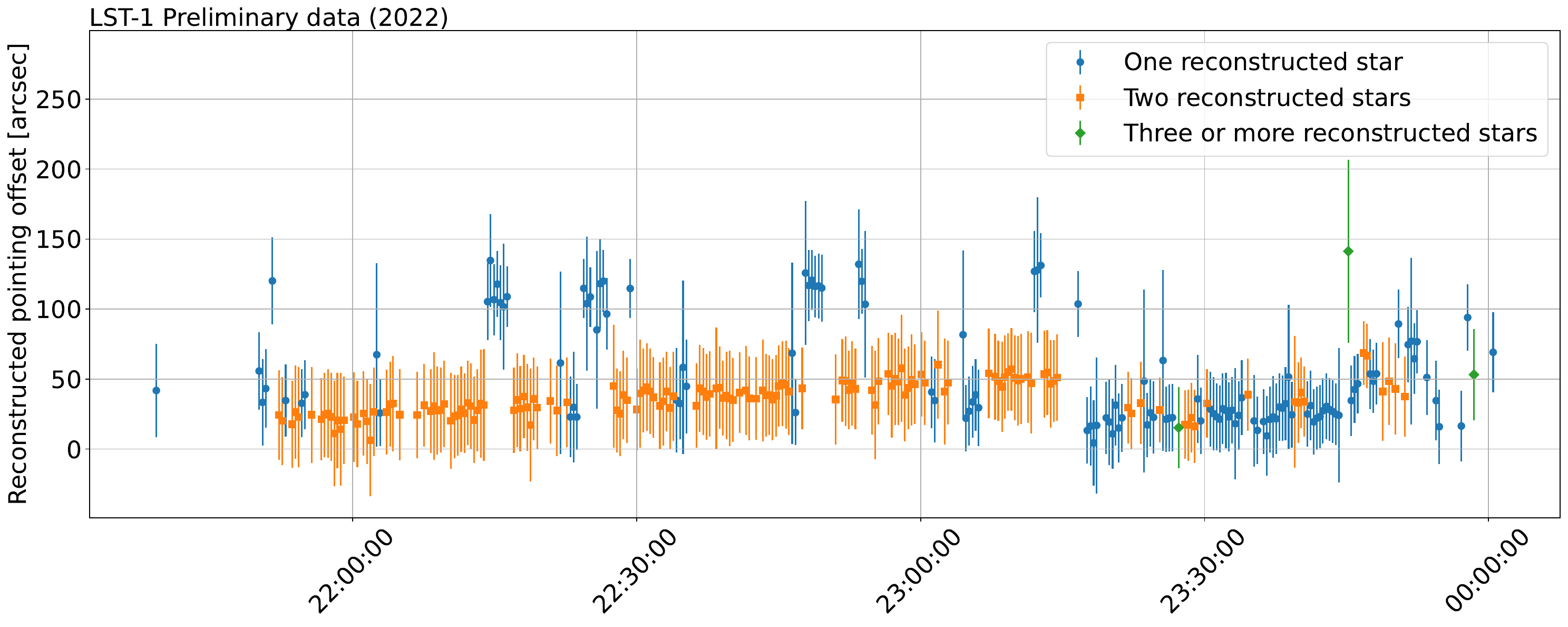}
	\caption{Reconstructed mispointing as a function of time, LST-1 data.}
	\label{fig:total_mispointing_vs_time_data}
\end{figure}

The simulation results, shown in Fig.~\ref{fig:alt_az_mispointing_vs_time}, demonstrated good accuracy in reconstructing the telescope's pointing offset, with better accuracy observed in the zenith distance direction compared to azimuth. The residual difference between the simulated and reconstructed pointing offset is attributed to the particular positions of stars in relation to the camera center and the spread of starlight over zenith distance due to the degrading point spread function (PSF) in the off-axis direction. The method achieved high accuracy below $15''$.

The star tracking method was then applied to real data taken by the LST-1 telescope.
Results, presented in Fig.~\ref{fig:total_mispointing_vs_time_data}, indicate that a single reconstructed star trajectory is insufficient for stable and unbiased pointing reconstruction, but with two or more reconstructed stars, the method demonstrates notable stability. The maximum mispointing observed is below $100''$, and mostly within the targeted tracking accuracy of $60''$. The telescope's bending model and final corrections to the active mirror control settings were still being refined at the time of data acquisition, potentially affecting the telescope's pointing accuracy, especially at small zenith angles. The telescope pointing appears stable except at very high elevation angles, with one notable outlier observed at around $8^\circ$\,zenith (around 23:45). This could be attributed to incorrect camera pixel calibration or an actual deviation of the telescope's drive from the intended trajectory. The analysis of interleaved calibration events is expected to improve the robustness of the single star reconstruction and help to eliminate such outliers. 

\acknowledgments
This work was conducted in the context of the CTA-LST Project.
We gratefully acknowledge financial support from the following agencies and organizations: \href{https://www.lst1.iac.es/acknowledgements.html}{CTA-LST Project Acknowledgements}.
\clearpage
\section*{Full Author List: CTA-LST Project}

\tiny{\noindent
K. Abe$^{1}$,
S. Abe$^{2}$,
A. Aguasca-Cabot$^{3}$,
I. Agudo$^{4}$,
N. Alvarez Crespo$^{5}$,
L. A. Antonelli$^{6}$,
C. Aramo$^{7}$,
A. Arbet-Engels$^{8}$,
C.  Arcaro$^{9}$,
M.  Artero$^{10}$
% ...
}\\

\tiny{\noindent
$^{1}$Department of Physics, Tokai University.
$^{2}$Institute for Cosmic Ray Research, University of Tokyo.
$^{3}$Departament de Física Quàntica i Astrofísica, Institut de Ciències del Cosmos, Universitat de Barcelona, IEEC-UB.
$^{4}$Instituto de Astrofísica de Andalucía-CSIC.
$^{5}$EMFTEL department and IPARCOS, Universidad Complutense de Madrid.
$^{6}$INAF - Osservatorio Astronomico di Roma.
$^{7}$INFN Sezione di Napoli.
$^{8}$Max-Planck-Institut für Physik.
$^{9}$INFN Sezione di Padova and Università degli Studi di Padova.
$^{10}$Institut de Fisica d'Altes Energies (IFAE), The Barcelona Institute of Science and Technology.
% ...
}

\end{document}